\documentclass[utf8]{FrontiersinHarvard} 

\usepackage{url,hyperref,lineno,microtype,subcaption}
\usepackage{amssymb}
\usepackage[onehalfspacing]{setspace}
\usepackage{amsmath}
\usepackage{placeins}

\def\firstAuthorLast{Triggiani {et~al.}} 
\def\Authors{Francesca Triggiani\,$^{1,2}$, Tommaso Morresi\,$^{3,4}$, Simone Taioli\,$^{3,4,5*}$  and Stefano Simonucci\,$^{1,2*}$}
\def\Address{$^{1}$School of Science and Technology, University of Camerino, Italy \\
$^{2}$INFN, Sezione di Perugia, Italy \\
$^{3}$European Centre for Theoretical Studies in Nuclear Physics and Related Areas (ECT*), Bruno Kessler Foundation, Trento, Italy \\
$^{4}$Trento Institute for Fundamental Physics and Applications (TIFPA-INFN), Trento, Italy\\
$^{5}$Faculty of Applied Physics and Mathematics, Gdańsk University of Technology, Gdańsk, Poland}
\def\corrAuthor{Corresponding Authors}

\def\corrEmail{stefano.simonucci@unicam.it; taioli@ectstar.eu}

\begin{document}
\onecolumn
\firstpage{1}

\title[Electron-water elastic scattering]{Elastic scattering of electrons by water: an ab initio study} 

\author[\firstAuthorLast ]{\Authors} 
\address{} 
\correspondance{} 

\extraAuth{}

\maketitle

\begin{abstract}
In this work we devise a theoretical and computational method to compute the elastic scattering of electrons from a non-spherical potential, such as in the case of molecules and molecular aggregates. Its main feature is represented by the ability of calculating accurate wave functions for continuum states of polycentric systems via the solution of the Lippmann-Schwinger equation, including both the correlation effects and multi-scattering interference terms, typically neglected in widely used approaches, such as the Mott theory. Within this framework, we calculate the purely elastic scattering matrix elements. 
As a test case, we apply our scheme 
to the modelling of electron-water elastic scattering. The Dirac-Hartree-Fock self-consistent field method is used to determine the non-spherical molecular potential projected on a functional space spanned by Gaussian basis set. By adding a number of multi-centric radially-arranged $s$-type Gaussian functions, whose exponents are system-dependent and optimized to reproduce the properties of the continuum electron wave function in different energy regions, we are able to achieve unprecedented access to the description of the low energy range of the spectrum ($0.001< E < 10$ eV) up to keV, finding a good agreement with experimental data and previous theoretical results.
To show the potential of our approach, we also compute the total elastic scattering cross section of electrons impinging on clusters of water molecules and zundel cation. 
Our method can be extended to deal with inelastic scattering events and heavy-charged particles.

\end{abstract}

\section{Introduction}

A complete understanding of the collision processes of charged particles with atoms, molecules, and solids is of paramount importance in several areas of physics, chemistry, and medicine.  
On the one hand, electron and ion beams are used for manufacturing devices, such as in lithography techniques to fabricate integrated circuits or other nanostructures \citep{Dapor_2020}.
On the other hand,
scattering experiments involving charged projectiles may represent a tool to ``extend our vision''. Indeed, by tuning the kinetic energy of the impinging particles,
we can characterise the spectrum of materials electronic and optical properties and probe different time (or frequency) and length (or momentum) scales, such as in energy loss \citep{azzolini2017monte,azzolini2018anisotropic,azzolini2020comparison,pedrielli2021electronic,pedrielli2022search}, secondary electron \citep{azzolini2018secondary,taioli2022role}, and Auger \citep{taioli2009mixed,taioli2010electron,colle2004ab2,colle2004auger,taioli2015computational,taioli2021resonant,taioli2015tetrapeptide} spectroscopies.
Moreover, the modelling of charge transport and flow inside materials \citep{Dapor_2020} and different media, such as stellar and interstellar plasma \citep{maciel_2013} or biological matter \citep{PMID:27652826}, is fundamental for the interpretation of phenomena ranging from the nucleosynthesis of the chemical elements in evolved stars \citep{simonucci2013theoretical,taioli2022theoretical,palmerini2021presolar,mascali2022novel,vescovi2019effects,morresi2018nuclear}
to hadrontherapy treatments for cancer cure \citep{de2022energy,taioli2020relative,pedrielli2021electronic}.\\
\indent In all these processes the charged particles undergo scattering events that can be classified as either elastic or inelastic. 
In inelastic scattering energy is exchanged between the projectile and the target constituents resulting in energy loss of the incident particle. 
The loss pattern, which can be caused by several mechanisms, can be used to analyse the target properties. 
For example, in materials science electron beams are employed as a probe of the electronic and optical properties, such as in analytical techniques like scanning (SEM) and transmission (TEM) electron microscopy. The energy transferred to the specimen produces an energy loss of the primary beam, which leads to a range of useful signals that can be used to characterize the material.
At variance, elastic scattering occurs with no energy loss experienced by the incident primary electron. Elastically scattered electrons change their direction of motion without changing their wavelength. For example, coherent elastic scattering is a common mechanism exploited in electron diffraction to analyze the structure of bulk crystals and surfaces, such as in the low energy electron diffraction (LEED) technique. \\
\indent Despite the recognized maturity of the experimental research field in electron scattering and charge transport, the quality (or nature) and the quantity of the experimental results available, the theoretical assessment of the cross sections of scattering events involving electrons is still essential for enabling us to resolve the combined effects on the spectral lineshape due to the intrinsic features of the intraband and interband transitions, the vibrational and rotational details, and the
characteristics of the incident beams and of the electron spectrometer apparatus. 
To deliver such an accurate theoretical description of the irradiation and of its role to initiate physical-chemical processes, one must develop a framework
capable of dealing with three main issues. First, a treatment of the electron repulsion that includes the correlation effects within the system and between the projectile and the (ionic) system, particularly at low energy where these effects are more significant;
second, the calculation of the continuum orbital at
a given positive energy, as the scattering matrix is rather sensitive to the quality of such wave function; finally, a computational effort that should be of the same order of magnitude as that of standard bound state calculations.\\ 
\indent In this work we focus on these three 
aspects by developing a theoretical and computational method for modelling the elastic collision of electrons from non-spherical potentials, which we apply as a test case to electron-water molecules scattering, a well-known and much studied problem. Water is indeed the main constituent of biological tissues, and electron-water collisions are  crucial events e.g. in the hadrontherapy techniques for cancer cure that aim to induce unrecoverable bio-damage to tumor cells. Surprisingly, the radiation damage is mostly due to low-energy electrons, generated upon collisions of impinging hadrons, such as protons or charged ions, with biological tissue rich in water \citep{gorfinkiel2005electron,taioli2006wave,taioli2006waterwaves,de2022energy,taioli2020relative}. 
However, the mechanisms underlying the radiation-based killing of tumor cells in cancer therapies have not yet been fully understood, especially at low energy where quantum effects are dominant \citep{song_2021,signorell_2020}. In fact, in our analysis we were driven by the lack of reliable experimental and theoretical data in low-energy (below 10 eV) electron-water elastic scattering, despite the study of electron scattering in liquid water being a very active field of research \citep{signorell_2020,gianturco_1998,gianturco_1987,itikawa_2005,gorfinkiel_2002,zhang_2009,faure_2004,kadokura_2019,gorfinkiel2005electron,taioli2006wave,song_2021}. Theoretical analysis is in particular hindered by the intrinsic difficulty to fully account for the physical properties of water assemblies from first principles, such as the existence of a permanent dipole moment, the electron-electron exchange-correlation interaction between the incoming electron and the scattering center, the numerous electronic excited configurations coupled with roto-vibrational states, and the inter- and intra- molecular multiple scattering effects.\\ 
\indent Indeed, it was immediately understood that the elastic scattering cross section for a polar molecule diverges at all impinging electron energies within the Born approximation, where nuclei are fixed \citep{altshuler_1957,garrett_1972} (as in the case of a bare Coulomb potential generated by a point-particle, see e.g. the Rutherford elastic cross section), and that rotations should be taken into account to achieve a finite value. Moreover, it was also realised that the electron elastic scattering is 
dominated by small scattering
angles \citep{fabrikant_2016, fedus_2017}.
In fact, also in currently and widely used approaches based on the Mott's theory \citep{salvat_2005} these effects are included semi-empirically, while the inter-atomic potential is treated as a mere superposition of atomic contributions neglecting the overlapping tails of the atomic potentials. More recently, Tennyson and coworkers 
used the R-matrix approach to study the electron and positron collisions with polar molecules, such as water, at low energy \citep{gorfinkiel_2002,zhang_2009,faure_2004,kadokura_2019,gorfinkiel2005electron,taioli2006wave,song_2021}. The R-matrix approach to scattering consists in dividing the system into an asymptotic region, where the non-interacting wave function is analytically known, and a scattering region, where the electron-molecule interaction is accounted for by high-accuracy quantum-mechanical methods, such as configuration interaction or coupled cluster. The matching of the inner and outer solutions at the boundary leads to important scattering information, such as phase shift and transmission probability. Nevertheless, these analyses were related to a single water molecule for impinging electron kinetic energies larger than 10 eV, also due to the lack of experimental data, typically carried out in water vapour, in this energy range.\\
\indent With this in mind, our primary goal is to develop a method that does not suffer the limitations of the previous approaches and allows one to reproduce accurately the experimental
results; moreover it is sufficiently general to be applied not only to
the analysis and interpretation of the elastic scattering cross sections of electrons colliding with (polar) 
 molecular aggregates but also
to be extended to the study of other inelastic scattering processes, such as excitation and ionization, since its central feature is the ability
of calculating accurate wave functions for continuum states of polycentric systems. \\
\indent Our approach
to assess the elastic scattering cross section (ESCS) of electrons (in general applicable to charged particles) is fully ab initio and consists in the following steps. First, we solve self-consistently the Dirac-Hartree-Fock (DHF) equations for the single water molecule or the water clusters, thus taking into account relativistic effects that may include spin-flip and polarization of the beam spin. 
We expand the electronic wave functions and the Coulomb potential for fixed nuclear positions using an Hermite Gaussian function (HGF) basis set, which is enlarged and optimized with more or less diffuse functions to reproduce the properties of the continuum electron orbital in different energy regions.
We then build the scattering cross section by computing the solution of the Lippmann-Schwinger (LS) equation for the scattering state \citep{taioli2010electron} with the Coulomb potential term projected into the functional space spanned by the Gaussian basis set.\\
\indent At first, we benchmark our method by comparing the results obtained using finite differences with one- and two-dimensional model potentials, where exact solutions for the scattering phases can be computed by accurately integrating the Schroedinger (or Dirac) equations. 
Moreover, we discuss elastic scattering on a single water molecule, and we further extend the investigation to clusters of water molecules in liquid phase and water aggregates, such as the zundel cation, to mimic the presence of surrounding molecular environment in order to rigorously determine the effect of multiple scattering on the elastic scattering cross section. 
\\
\indent 
This article is organised as follows:
in Section 2 we will discuss the basics of our theoretical and computational method, such as the projected potential method using HGF basis set in connection with the LS equation,
and its benchmark with a toy model potential in one and two dimensions; to show its capabilities and accuracy, in Section 3 we apply our framework to the calculation of the total 
elastic cross section of electrons colliding with a single, a cluster, and an aggregate of water molecules. 
In Section 4 we present final remarks and conclusions. 

\section{Theoretical and computational method}

\subsection{Differential and total elastic scattering cross sections}

The ESCS, which we aim to compute in this work, is defined as follows: 
\begin{equation}\label{total_cs}
\sigma_{el}(T)=\int_{\Omega}\frac{d\sigma_{el}(T,\Omega)}{d\Omega}d\Omega,
\end{equation}
where $T$ is the kinetic energy of the primary beam, and $d\Omega=2\pi \sin\theta d\theta d\phi$ defines the differential solid angle.
The differential elastic scattering 
cross section (DESCS) appearing in Eq. \ref{total_cs} can be obtained by following two different routes. \\
\indent On the one hand one can apply the 
relativistic Mott theory \citep{mott_1929}, whereby the DESCS of electrons impinging on a single molecule can be computed as \citep{Dapor_2020,salvat_2005}:

\begin{equation}\label{molsca}
\frac{d\sigma_{\mathrm{el}}(T,\theta)}{d\Omega}\,={\mathcal N}\,\sum_{m,n}\,e^{(i {\bf q} \cdot {\bf r}_{mn})}\,[f_m(\theta) f^*_n(\theta)\,+\,g_m(\theta) g^*_n(\theta)]\,,
\end{equation}
where ${\bf r}_{mn}\,=\,{\bf r}_m\,-\,{\bf r}_n$ is the distance between the atoms $m$ and $n$ within the molecule, ${\mathcal N}$ is the target atomic number density, and ${\bf q}$ and $\theta$ are the momentum transfer and the scattering angle; finally $f_{m,n}(\theta)$ and $g_{m,n}(\theta)$ are the direct and spin-flip scattering amplitudes of the atomic species $m,n$, respectively. $f_{m,n}(\theta)$ and $g_{m,n}(\theta)$ can be obtained by solving the Dirac equation in a central atomic potential for the single atomic constituent.\\
\indent To take into account the presence of bonded interactions between different chemical species in Eq. \ref{molsca} the sum is carried out using amplitudes weighted by an exponential displacement rather than probabilities, allowing for interference and intramolecular scattering. Nevertheless, within this approach the calculation of the molecular scattering potential and of the continuum electron wave function is carried out as a mere superposition of independent atomic centers, thus neglecting the overlapping tails of the long-range part of the Coulomb interaction and the relevant multiple scattering effects. In this framework, consequently, the presence of the electrical polarity due to charge separation, such as the permanent dipole moment of liquid water that has a dramatic effect on the DESCS, is superimposed and treated semi-empirically. \\
\indent The presence of surrounding water molecules in the liquid phase, which are randomly oriented can be taken into account by averaging the Eq. \ref{molsca} over all the possible spatial orientations, which leads to the following relation:
\begin{equation}\label{eq:Mott}
\frac{d\sigma_{el}}{d\Omega}=\sum_{m,n} \frac{\text{sin} \left(k r_{m,n} \right)}{kr_{m,n}} (f^*_{m}(\theta) f_{n}(\theta)+g^*_{m}(\theta)g_{n}(\theta)).
\end{equation} 
Again, Eq. \ref{eq:Mott} does neglect both multiple scattering effects, as atoms are treated as independent scattering centers, and the overlap between the long-range parts of the intermolecular potential.\\ 
\indent
To overcome these issues, we have developed an approach based on the full ab initio treatment of the electron-electron (intermolecular and intramolecular) interaction.
Within our approach, the DESCS is assessed by: 
\begin{equation}\label{eq:diff_elastic_cross_sect}
\frac{d\sigma_{el}}{d\Omega}=\frac{m^2}{4\pi^2}|\langle{\phi_{k\hat n}}|T^+(E)|\phi_{\mathbf{k}}\rangle|^2=\frac{m^2}{4\pi^2}|\langle{\phi_{k\hat n}}|V|{\psi^+_{\mathbf{k}}}\rangle|^2,
\end{equation} 
where $m$ is the mass of the impinging particle, $\phi_{k\hat n}$ and $\phi_{\mathbf{k}}$ are the asymptotic electron wave-function long after and long before the scattering, respectively, characterised by the modulus of the momentum $k$, which is scattering invariant (elastic scattering), the scattering angle identified by the unit vector $\hat n$, and incident orientation defined by $\hat{\mathbf{k}}$; $T$ is the on-shell $T$-operator, $V$ is the self-consistent potential and $\psi^{+}_{\mathbf{k}}(E)$ is the scattering wave function.\\ 
\indent The latter quantity can be obtained by following two different, while equivalent routes. The first approach is the direct solution of the many-body Schroedinger (or Dirac in a fully relativistic treatment of the electron motion) hamiltonian for positive energy eigenvalues $E$:
\begin{equation}\label{hamilton}
    H\psi^{+}_{\mathbf{k}}(E)=(H_0+V)\psi^{+}_{\mathbf{k}}(E)=E \psi^{+}_{\mathbf{k}}(E),
\end{equation}
where $H_0$ and $V$ are the kinetic energy and the electron-electron Coulomb repulsion operators, respectively.
Furthermore, it has been proved \citep{taioli2010electron}
that such many-body problem can be reduced to an effective single-particle problem,
in which the scattering states can be obtained from the resolution of the following LS equation for the continuum orbital that can be solved exactly using a Green's
function approach with the proper boundary condition:
\begin{equation}\label{LS}
\psi^{+}_{\mathbf{k}}(E)=\phi^{+}_{\mathbf{k}}(E)+G_0^+(E)V\psi^{+}_{\mathbf{k}},
\end{equation}
where
\begin{equation}
G_0^+(E)=\lim_{\epsilon \rightarrow 0} (E-H_0+i\epsilon)^{-1}
\end{equation}
is the Green's function of the free particle at the same energy $E$ of the interacting hamiltonian in Eq. \ref{hamilton}, and $\phi^{+}_{\mathbf{k}}(E)$ is the eigenvector of the free hamiltonian $H_0$.
This expression holds in both non-relativistic and relativistic theories, notwithstanding wave functions are 4-components spinors in the latter framework \citep{reiher_2014}. In this work we adopt the Green's function method of Eq. \ref{LS} by projecting both the Coulomb potential and the
continuum orbital using HGFs (for details on the use of HGFs with the LS equation see the following section \ref{gauspproj}).  \\
\indent 
The DESCS in Eq. \ref{eq:diff_elastic_cross_sect} is computed by solving either the Schroedinger or the Dirac equation
within a numerical scheme where the electron-electron Coulomb repulsion is approximated. In our approach, we use a mean-field approximation based on the 
Hartree-Fock (or Dirac-Hartree-Fock (DHF) in a fully relativistic treatment) method, where the orbital wave function and the scattering potential are obtained by the self-consistent solution of the relevant equations (see the Appendix for further details). 

\subsection{Benchmark system: 1D multi channel scattering from toy model potentials}

Prior moving to the discussion of the results on water and water aggregates, as a test we
benchmark our method against a model system. In particular, we consider the following two-channels problem ($\hbar=1$):
\begin{equation}\label{eq:Schrodinger_2channels}
-\frac{1}{2m}\nabla^{2}\psi_{j}\left(\mathbf{r}\right)+\sum_{k}V_{jk}\left(\mathbf{r}\right)\psi_{k} \left(\mathbf{r}\right)=E \psi_{j}\left(\mathbf{r}\right),
\end{equation}
where $j,k$ label the indices of the two channels. The hamiltonian in Eq. \ref{eq:Schrodinger_2channels} is intentionally non-relativistic avoiding thus the description of the electrons in terms of spinors and total angular momentum, as our goal is to resort to the simplest model in order to test the validity of the LS approach in connection with HGFs as a basis set to expand continuum scattering states. In fact, in a non-relativistic approximation to scattering we rely on the use of wave functions and a good quantum number is represented by the orbital angular momentum. We also limit our test for the sake of simplicity to the modelling of two $s$-orbitals channels, as the additional basis functions that we include in the simulation of the electron-water scattering contain only $s$-type HGFs. \\
\indent 
By considering a spherically symmetric potential $V(r)$, the wave function can be conveniently factorised in radial and angular parts, i.e.
$\psi_{j}\left(\bf{r}\right)=\mathcal{R}_{jl}\left(r\right)Y_{lm}\left(\vartheta,\varphi\right)$, which specifically for the $s$-channels case reads $\psi_{j}\left(\bf{r}\right)=\mathcal{R}_{j0}/\sqrt{4\pi}$. From Eq. \ref{eq:Schrodinger_2channels} one writes the following system of coupled differential equations:
\begin{equation}\label{eq:rad_2channels}
\begin{split}
& -\frac{1}{2m}\frac{d^{2}}{dr^{2}}u_{1}\left(r\right)+ V_{11}\left(r\right)u_{1}\left(r\right)+V_{12}\left(r\right)u_{2}\left(r\right)=E u_{1}\left(r\right) \\
& -\frac{1}{2m}\frac{d^{2}}{dr^{2}}u_{2}\left(r\right)+ V_{21}\left(r\right)u_{1}\left(r\right)+V_{22}\left(r\right)u_{2}\left(r\right)=E u_{2}\left(r\right),\end{split}
\end{equation}
where $u_{j}\left(r\right)=r\mathcal{R}_{j}\left(r\right)$, and $u:\mathbb{R}_{+}\rightarrow\mathbb{C}^{2}$
fulfills the condition  $u_{j}\left(0\right)=0$. We solve these equations by further assuming $m=1$ for the following model potentials:
\begin{equation}\label{RGLS}
\begin{split}
& V_{11}\left(r\right)=-V_{22}\left(r\right)=-14\frac{e^{-0.5r}}{r}+7\frac{e^{-0.2r}}{r} \\
& V_{12}\left(r\right)=V_{21}\left(r\right)=2\frac{e^{-0.3r}}{r}.
\end{split}
\end{equation}
\begin{figure}[hbt!]
\begin{center}
\includegraphics[scale=0.69]{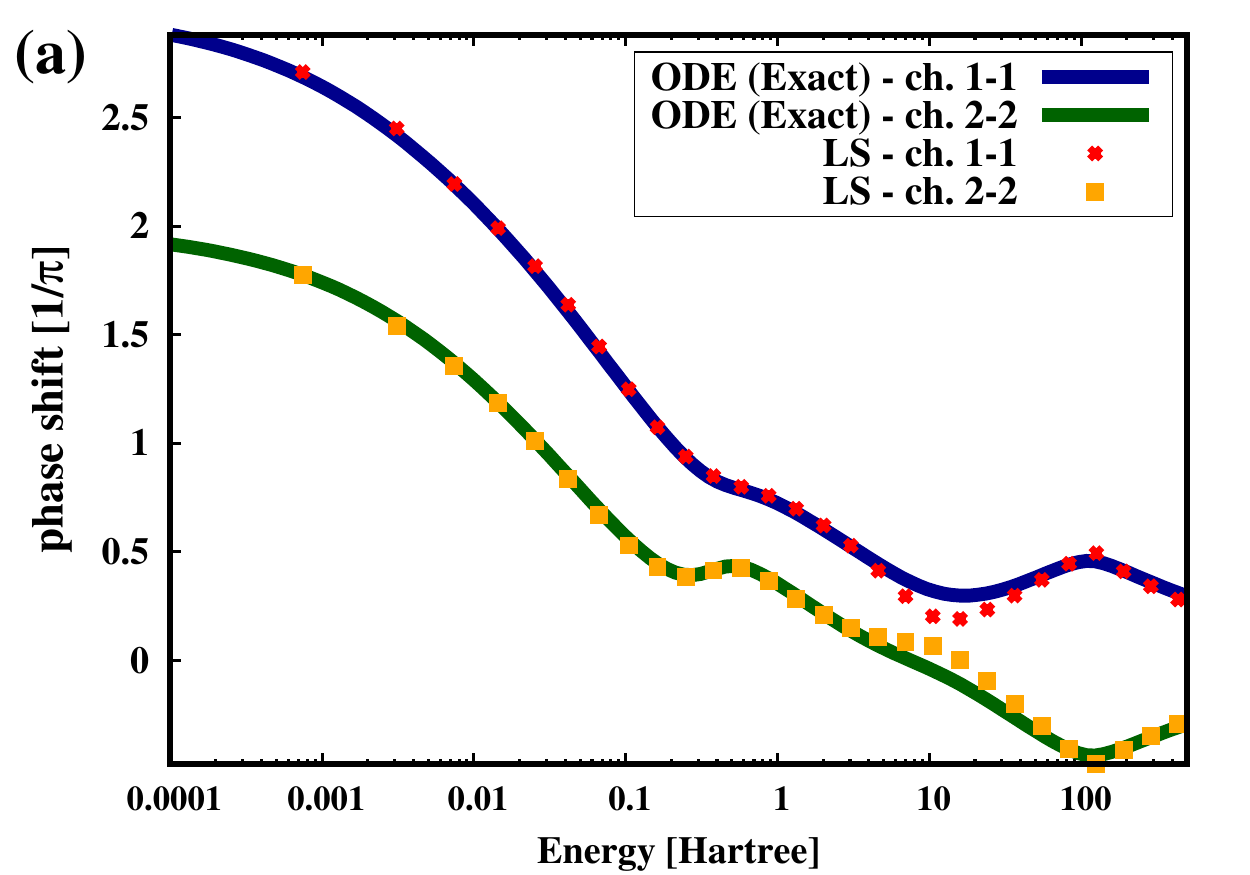}
\includegraphics[scale=0.69]{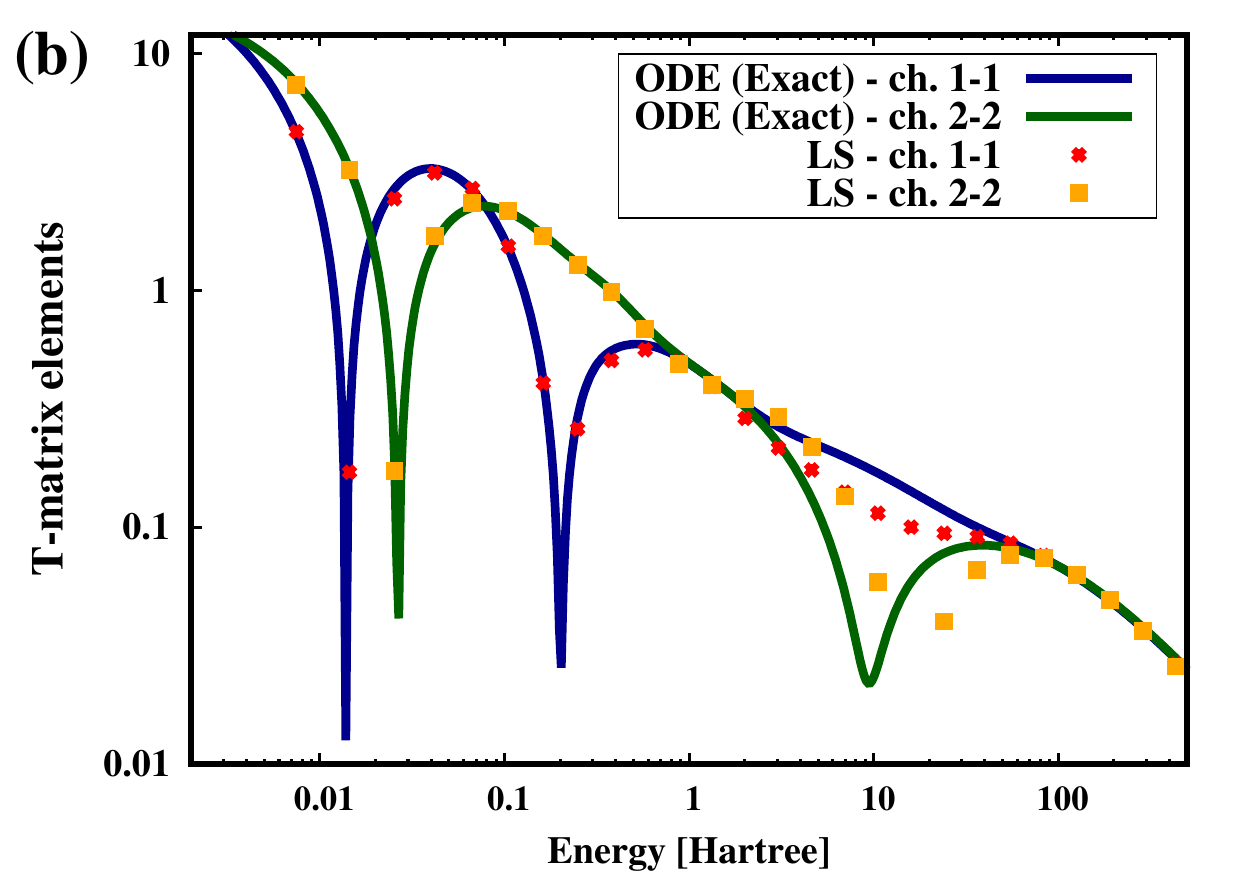}
\end{center}
\caption{Results of the toy model potential problem of Eq. (\ref{eq:rad_2channels}). (a) Phase shifts for the auto-channels obtained through the integration of the differential equation (ODE) and through the projection on a Gaussian basis set by solving the corresponding Lippmann-Schwinger equations. The Gaussian basis set is composed by 40 well tempered $s$-Gaussian functions. (b) Comparison of the $T$-matrix elements for the auto-channels using the LS and ODE approaches.}\label{fig:toy_model}
\end{figure}
In Figure \ref{fig:toy_model} we show the phase shifts (panel a) and the $T$-matrix elements (panel b) calculated in the auto-channels, identified as those channels in which both the incoming and outgoing wave funtions are in the same channel.\\
\indent In particular, we compare the numerical solution obtained by integrating the Eqs. \ref{eq:rad_2channels} via a fourth-order Runge-Kutta method (ODE in Figure \ref{fig:toy_model} represented by the colored continuous lines), which is almost exact, with the results obtained using the LS equation (points in Figure \ref{fig:toy_model}) in which the potential is projected using 40 well-tempered $s$-type Gaussian functions \citep{huzinaga_1993}, which deliver the best convergence with the ODE calculations (for details on the use of Gaussian functions with the LS equation see the following section \ref{gauspproj}).
We observe that the phase shifts in Figure \ref{fig:toy_model}a) in the auto-channels are well reproduced apart from a small region around 10 Hartree where the LS solutions deviate from the exact ones. This is reflected in the $T$-matrix elements of Figure \ref{fig:toy_model}b), where the resonance in the channel 2-2 around the energy of 10 Hartree is shifted towards larger energies in the LS solution. This can be rationalized by recognizing that the basis set is not well suited to describe accurately this energy region, while it works very well in the low energy part. It is worth to notice that, of course, the resolution achievable by a given basis set depends upon the number and type of Gaussian functions. We will solve this issue in the calculation of the ESCS of water.   

\subsection{Our method: Gaussian projection of the Lippmann-Schwinger equation}\label{gauspproj}

The numerical solution of the DHF equation (or Schroedinger equation in non-relativistic approximation) can be found by expanding the wave functions and the potential using HGFs, which represent one of the most suitable basis set to study non-spherical polyatomic systems, such as water, water cluster and water molecular aggregates (for more details on this procedure see Ref. \citep{taioli_2021}).
The HGFs are defined as follows:
\begin{equation}\label{gaussian}
g(\mathbf{r})=g(u,v,w;a,\mathbf{R};\mathbf{r})=N{\partial^{u+v+w}\over
\partial X^u \partial Y^v\partial Z^w}({2a\over\pi})^{3/4}
\exp[-a (\mathbf{r}-\mathbf{R})^2],
\end{equation}
where $\mathbf{R}\equiv(X,Y,Z)$ is the center of the HGF (which can or cannot coincide with the nuclei positions), $a$ determines the HGF width, ($u,v,w$) represent the order of derivation that determines the symmetry of the HGF (e.g. first derivative represents a $p$-type orbital, the second derivative a $d$-type orbital...); finally, $N$ is a coefficient set to normalize the HGFs.
The mono- and bi- electronic integrals of the fully relativistic many-body hamiltonian are evaluated by including several HGFs characterised by different derivation orders and centers \citep{TAIOLI2010237,taioli_2021}.\\
\indent Our approach relies on the semi-projected potential method \citep{taioli2010electron,taioli_2021}, where the Coulomb potential is projected in HGFs, while the kinetic energy term $H_0=c{\bf \alpha} \cdot {\bf p} +\beta mc^2$ ($\alpha$ and $\beta$ are the Dirac matrices, see Eq. \ref{alpha} of the Appendix) is unprojected so as to recover the asymptotic continuum after the Coulomb potential dies off.
Within this method the LS equation \ref{LS} for the scattering wave function identified by the quantum number $\gamma$ reads:
\begin{equation}\label{LSp1}
\psi_{\gamma {\bf k}}^+({\bf r})=e^{i \bf{k} \cdot \bf{r}}+
  G_0^+(\varepsilon) V_\gamma^t(\bf {r})\psi_{\gamma \bf{k}}^+(\bf{r}),
 \end{equation}
where $V_\gamma^t=\sum_{\lambda\mu\nu\tau}
  |\lambda>S^{-1}_{\lambda\mu}<\mu|\hat V_\gamma|\nu>S^{-1}_{\nu\tau}
  <\tau|,\quad S_{\lambda\mu}=<\lambda|\mu>$ is the approximate DHF self-consistent potential represented by a set of HGFs, which includes both the
Coulomb and static dipole components of the potential. 
By replacing $ V_\gamma$ with $V^t_\gamma$ the LS equation \ref{LSp1} can be solved as follows:
\begin{equation}
\psi_{\gamma\bf{k}}^+(\vec r)=e^{i\bf{k}\cdot\bf{r}}+ G_0^-(\varepsilon)T_\gamma(\varepsilon)e^{i\bf{k}\cdot\bf{r}},
\end{equation}
where $T_\gamma$ is the transition operator defined by the equation $T_\gamma= V^t_\gamma+ V^t_\gamma G_0^-(\varepsilon) T_\gamma$.
Within this approach we make use of the
{\it Static Exchange Approximation}, in which one neglects the effects of the continuum orbital on the bound orbitals.
The elements of the HGF basis set are chosen to minimize the difference
$(V_\gamma-V^t_\gamma)\psi_{\gamma \bf{k}}^+(\bf{r})$ inside the region of interest. \\
\indent As in this work we aim to assess the total elastic scattering cross section for a range of electron kinetic energies, which can span several orders of magnitudes (from 0.001 to 10$^3$ eV), one can guess that it is cumbersome to find a unique HGF basis set able to reproduce the behaviour of the scattering wave function in such a large energy region.
In particular, we have found that the basis set for different energy ranges cannot be one and the same, owing to the difficulty of reproducing the continuum orbital at different regimes with the same basis set. In our simulations of the liquid water we use thus two sets of HGFs.
The first basis set is centered on the oxygen and hydrogen atoms and is common to all the test cases, independent of the number of water molecules included in the calculation of the elastic scattering cross section.
In particular, the basis sets used to expand the bound and continuum orbitals are made of totally uncontracted HGFs, that is of $(12s,8p,6d,4f)$ Gaussian functions centered on each hydrogen nucleus and of $(20s,12p,8d,6f,4g)$ Gaussian functions centered on the oxygen nuclei. These uncontracted basis functions must be optimized so as to reproduce orbital eigenfunctions of the DHF hamiltonian at the energies of interest. \\
\indent Depending on the electron kinetic energy range, we have found that the inclusion of a second set of HGFs is critical to reproduce the elastic scattering experimental data. This observation is in agreement with the toy model potential calculations previously presented, where a small number of HGFs was not sufficient to reproduce the accuracy of the exact solution at all energies (see Figure \ref{fig:toy_model} around 10 Hartree).
This finding can be rationalized by noticing that the insertion of this further $s$-symmetry basis set delivers enough flexibility to reproduce the scattering wave function behaviour in different energy regions: narrower HGFs (that is larger exponential coefficients $a$ in Eq. \ref{gaussian}) can more accurately represent the highly oscillating behaviour of the scattering wave function at high energy, while more diffuse HGF (that is smaller exponential coefficients $a$ in Eq. \ref{gaussian}) can better reproduce the slowly varying behaviour of the scattering wave function at low energy.
This further HGF basis set is uniformly distributed within a sphere, roughly centered on the center of mass of the water molecule, and characterised by a variable radius $R=1.25/\sqrt{a}$ (in a.u.), where $a$ is the exponent of the HGF.

\section{Results and discussion}

\subsection{Single water molecule}

\begin{figure}[htbp!]
\begin{center}
\includegraphics[width=0.99\textwidth,trim={0cm 0.6cm 0cm 0cm},clip=true]{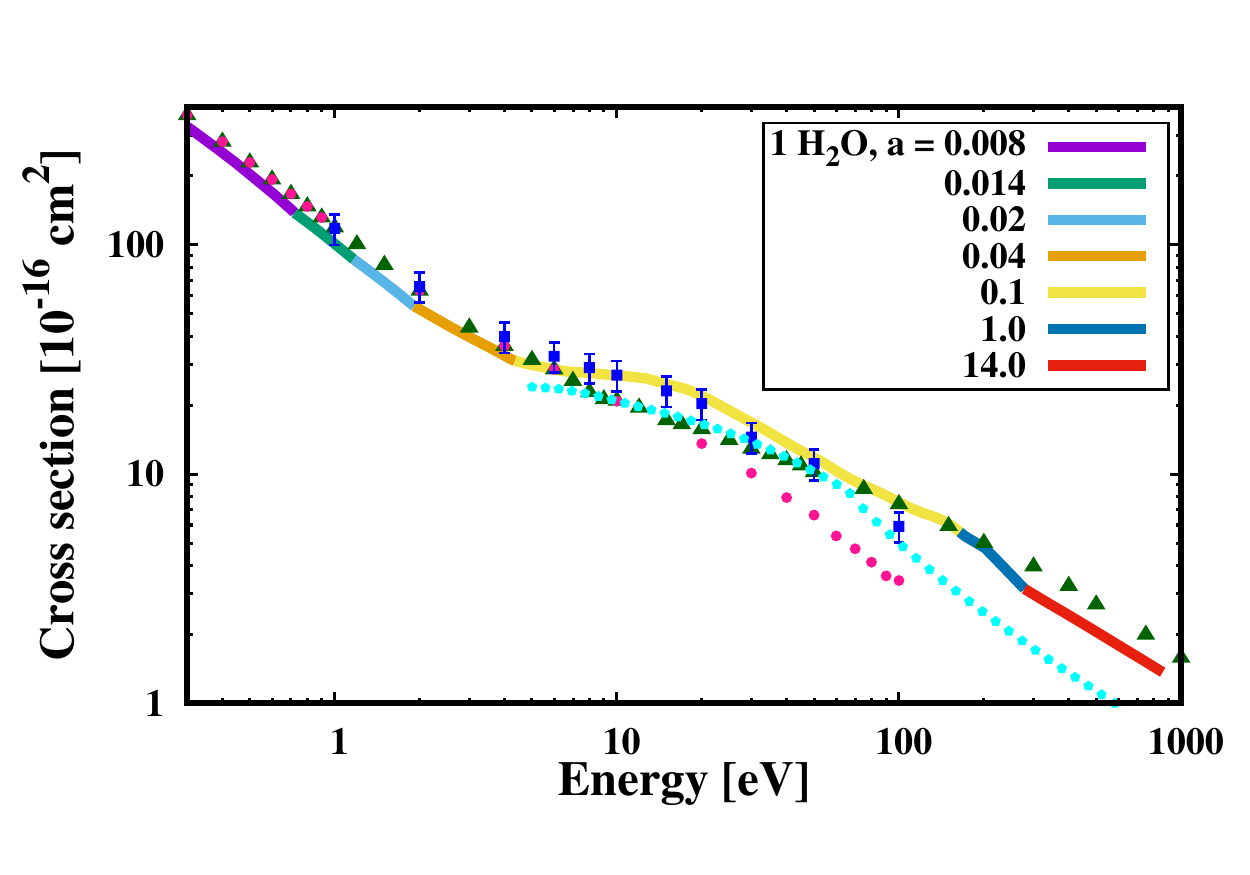}
\end{center}
\caption{Theoretical elastic scattering  cross section (continuous multi-colored line) of electrons impinging on a single water molecule with different kinetic energies compared to mixed experimental/theoretical results from Ref. \citep{khakoo_2013} (full blue squares with error bars), with the recommended elastic (full purple circles) and total (green triangles) cross sections from Ref. \citep{song_2021}. The recommended elastic cross section of Ref. \citep{song_2021} is based on the results given in Ref. \citep{gorfinkiel_2002} up to 6 eV and by interpolating the cross sections obtained with different methods \citep{itikawa_2005}. The recommended total cross section is obtained at energies between 0.1 and 7 eV via R-matrix calculations \citep{zhang_2009,faure_2004}, at 7–50 eV is based on experiments from Refs. \citep{szmytkowski_1987,szmytkowski_2006,kadokura_2019} and above 50 eV on experimental measurements from Ref. \citep{munoz_2007}. The cyan circles report the results obtained by using the Elsepa code \citep{salvat_2005}. Data in the $x$ and $y$ axes are reported in $\log$ scale, $a$ is in bohr$^{-2}$.}\label{fig:single_molecule}
\end{figure}
Using the approach presented in the methodological section, we carried out the calculation of the total elastic scattering cross section of electrons impinging on i) a single water molecule; ii) a cluster of 2 and 3 water molecules; iii) the zundel cation. \\ 
\indent In Figure \ref{fig:single_molecule} we show the results obtained for the single water molecule, which is well studied both experimentally and theoretically. 
Our ab initio results (continuous multi-colored line) are compared with other theoretical and experimental approaches summarized in Ref. \citep{song_2021} (see the figure caption for details on the different points). We notice that our calculations are performed in the fixed-nuclei approximation, i.e. we do not account for rotational and vibrational degrees of freedom. We also stress that unfortunately a single set of HGFs cannot accurately describe the elastic cross section in the full spectrum, that is to say that one needs to include more or less diffuse HGFs in order to be able to describe the oscillations of the continuum wave function (and, ultimately, of the Coulomb potential tail) in different energy ranges.
Thus, we decided to add further $s-$symmetry HGFs to the bound state basis set of water, whose width $a$ depends on the energy region. Each colour of the line represents thus a different Gaussian width (see the legend in Figure \ref{fig:single_molecule}, reporting in bohr$^{-2}$ the seven $a$ coefficients used to accurately reproduce the experimental data).
We notice that while these coefficients where optimized by comparison with experimental results, their physical meaning is clear: at low energies the variance of the Gaussian functions should capture the long wavelengths related to low energies; on the other hand at higher energies the dependence on the $a$ coefficient is less stringent; in general, larger values of $a$ describe better the higher oscillations.
Interestingly, above 4 eV our results are in excellent agreement with those by Khakoo et al. \citep{khakoo_2013}, while below 4 eV our data, although following the same trend, are slightly outside the error bars. This is in accordance with the fact that the results of Ref. \citep{khakoo_2013} were obtained using the Schwinger multichannel method \citep{taketsuka_1981}, employing extensive basis sets and considering both polarization and dipole-scattering effects. In our DHF framework, of course we also take into account them, and our LS solutions must be stationary points of the Schwinger variational principle.
We also notice that at energies $>$ 10 eV, both our and Ref. \citep{khakoo_2013} results fit better the total cross section trend (green triangles in Figure \ref{fig:single_molecule}a)), than the recommended elastic one (full purple circles \citep{song_2021}). This point can be understood by noticing that our approach is based on the inclusion in our model of the elastic channel only, while the total cross section contains the contributions of both elastic and inelastic scattering events. However, we have previously demonstrated (see Refs. \citep{taioli2010electron,taioli2015computational}) that the total probability 
is not significantly modified by the opening of multiple channels; rather such total transition probability is redistributed among several channels, while keeping constant the total value. 
Thus, it is not surprising that our ESCS actually follows closely the experimental data. \\
\indent Finally, in Figure \ref{fig:single_molecule} we also compare our ab initio calculations with the Elsepa \citep{salvat_2005} model (cyan points), which is fully relativistic but the atomic rather than the molecular interaction potential is considered, while the dipole is included semi-empirically. The discrepancy between our and Elsepa results can be thus attributed to the different treatment of the interaction potential.


\subsection{Water clusters}

\begin{figure}[hbt!]
\begin{center}
\includegraphics[scale=0.6]{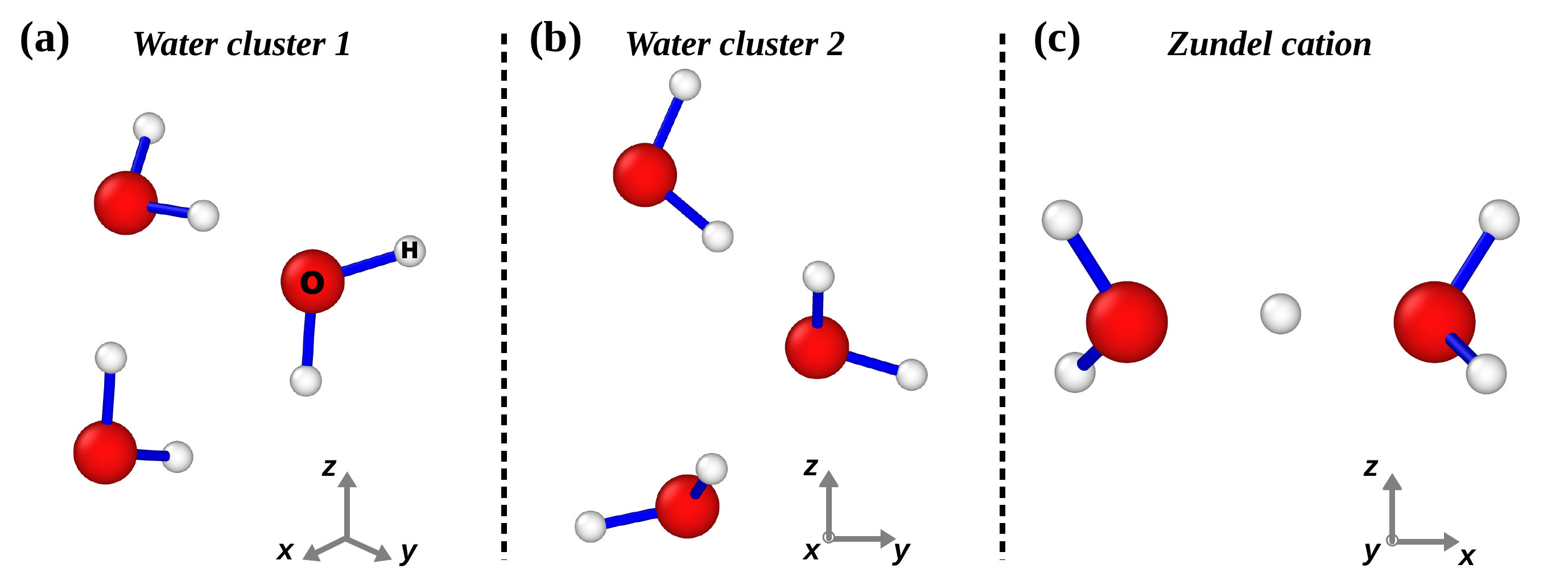}
\end{center}
\caption{Water clusters employed in this work as scattering centers. In panels (a) and (b) we report two different configurations of water molecules in the liquid phase. In panel (c) we show the zundel cation minimum configuration as reported in \citep{huang_2005}. 
}\label{fig:clusters}
\end{figure}

To show the potential of our method to deal with polycentric systems, we have studied the electron scattering with small clusters of water molecules, which are shown in Figure \ref{fig:clusters}.
In particular, in the panels a) and b) of Figure \ref{fig:clusters} we report two different clusters made of three molecules, which were obtained by trimming a large cubic cell of liquid water (i.e. density is 1 g / cm$^3$), whose configuration was optimized using Density Functional Theory implemented with a PBE functional \citep{qe1}. The two clusters
are meant to describe the different environment in which electrons travel inside liquid water, having the dipole moments pointing towards different directions.
\begin{figure}[htbp!]
\begin{center}
\includegraphics[width=0.99\textwidth,trim={0cm 0.6cm 0cm 0cm},clip=true]{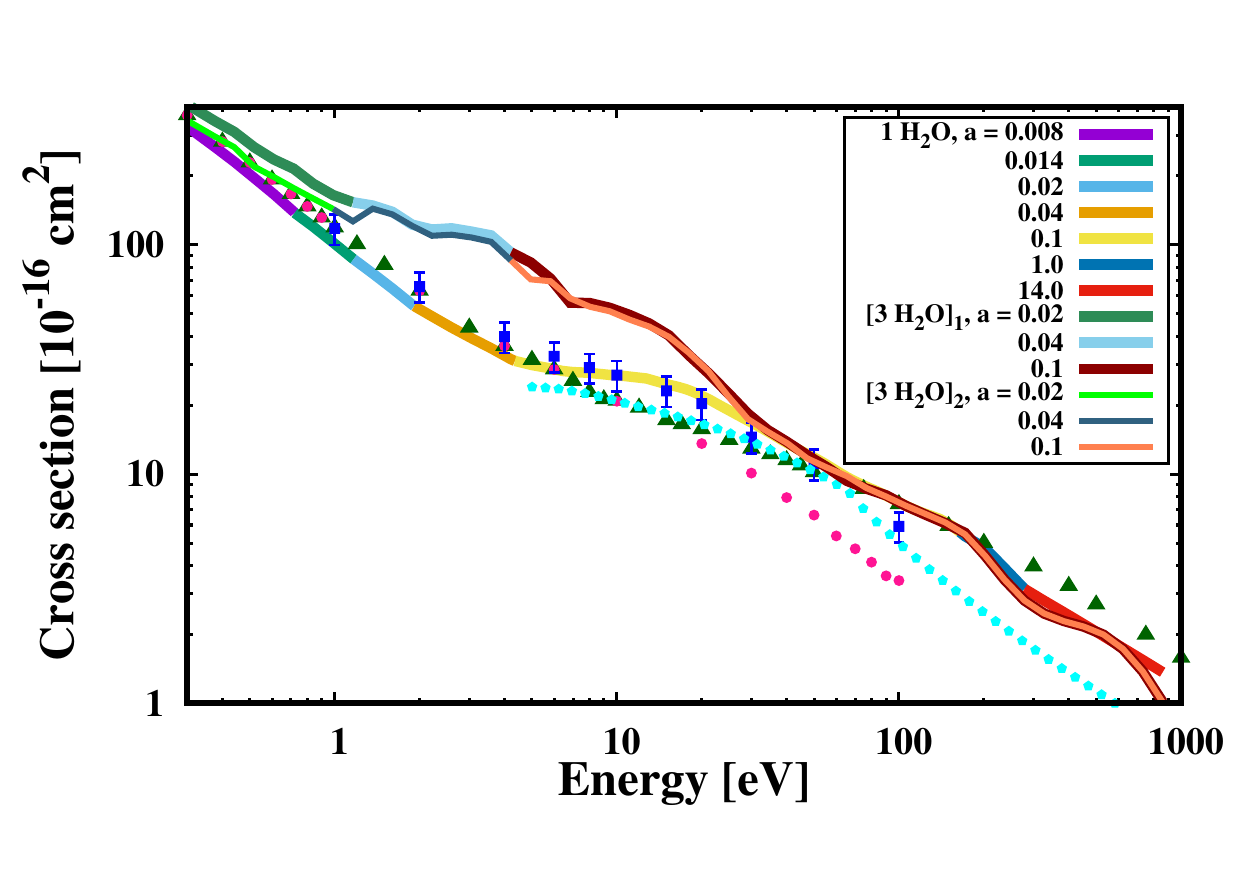}
\end{center}
\caption{ESCS of electrons colliding at different kinetic energies with the two water clusters (3 [H$_2$O]$_{1,2}$) reported in Figure \ref{fig:clusters}, in comparison to the results for the single water molecule (1 H$_2$O) shown in Figure \ref{fig:single_molecule}. Multi-colored lines report our calculations; the cyan points report the values computed with the Elsepa code \citep{salvat_2005} on the  water cluster 1 of Figure \ref{fig:clusters}. The total cross section of the water clusters has been divided by three, i.e. to normalize it with the number of molecules within the clusters. Data in the $x$ and $y$ axes are reported in $\log$ scale, $a$ is in bohr$^{-2}$.}\label{fig:cs_water_mol}
\end{figure}
In Figure \ref{fig:cs_water_mol} we show the ESCS of these two water clusters resulting from our simulations (upper multi-colored lines) in comparison to both experimental data (symbols), single water molecule calculations (lower multi-colored line), and Elsepa (cyan points). 
Each colour of the line represents a different Gaussian width (see the legend in Figure \ref{fig:cs_water_mol}, reporting in bohr$^{-2}$ the $a$ coefficients used in different energy ranges).
Of course, the ESCSs of the water clusters have been normalized by the number of molecules in the simulation cell. We notice that the ESCS of the two water clusters characterised by different water molecule configurations are similar. However, by comparing our simulations of the single water molecule with water clusters, we find a very good match in the low ($E<$ 1 eV) and high ($E>$ 30 eV) energy range; at odds, we obtain a diverging behaviour in the intermediate regime ($1<E<30$ eV) with ESCS differing by almost 100\% at around 3 eV. As the basis set we use is comparable in both size and diffuseness for the three cases investigated, we attribute such a discrepancy in the ESCS at intermediate energy due to multiple scattering and a more accurate representation of the inter- and intra-molecular potential. Indeed, Elsepa simulations of ESCS on cluster 1 of Figure \ref{fig:clusters} follow closely the single water molecule behaviour.
We stress that experimental measurements are carried out on water vapour in different conditions of relative humidity, while our simulations are carried out using the standard density of liquid water. This finding highlights the importance of an accurate treatment of the interatomic potential for dealing with liquid phases as compared to vapour phase, where molecules are far apart and, thus, more closely resembling the single molecule calculations. 

\subsection{Elastic scattering from the zundel cation}

Finally, to show the flexibility of our approach to deal with charged system we calculate the ESCS of electrons impinging on protonated water (zundel cation), in which hydrogen ions bond with water molecules and form loosely bound structures (see panel c) of Figure \ref{fig:clusters}).
\begin{figure}[htbp!]
\begin{center}
\includegraphics[width=0.99\textwidth,trim={0cm 0.6cm 0cm 0cm},clip=true]{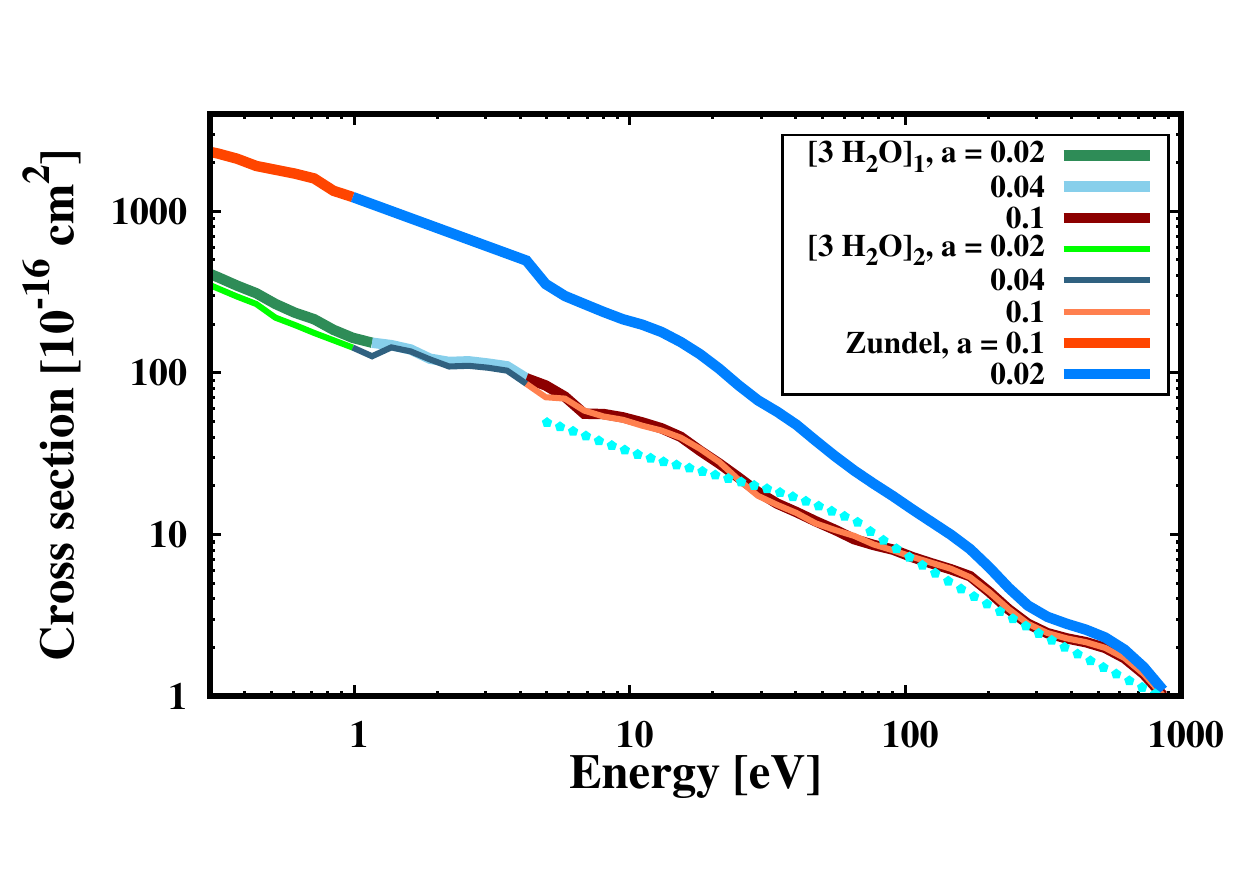}
\end{center}
\caption{ESCS of electrons colliding at different kinetic energies with the zundel cation in comparison to the water clusters (3 [H$_2$O]$_{1,2}$) reported in Figure \ref{fig:clusters}. Multi-colored lines report our calculations; the cyan points report the values computed with the Elsepa code \citep{salvat_2005} on the  water cluster 1 of Figure \ref{fig:clusters}. The total cross section of the water clusters has been divided by three, i.e. to normalize it with the number of molecules within the clusters, and by two for the zundel cation. Data in the $x$ and $y$ axes are reported in $\log$ scale, $a$ is in bohr$^{-2}$.}\label{fig:cs_zundel}
\end{figure}
In particular, in Figure \ref{fig:cs_zundel} we compare the results for the zundel cation (upper multi-colored line) obtained using our relativistic ab initio method with those of the water clusters (of course all ESCSs have been normalized with respect to the number of water molecules within the systems). Each colour of the line represents a different Gaussian width (see the legend in Figure \ref{fig:cs_zundel}, reporting in bohr$^{-2}$ the $a$ coefficients used in different energy ranges). We notice that the presence of different environments dramatically changes the ESCS, particularly in the intermediate to low energy region ($<100$ eV). This again shows up the importance of accurately modelling the intra- and inter- molecular potential, as the different behaviour of the curves in Figure \ref{fig:cs_zundel} can be attributed to the different environment and multiple scattering effects more than to a deficiency of the HGF basis set. Finally, we also report the Elsepa simulations carried out for the zundel cation configuration, even though we stress that Elsepa (cyan points) can deal only with neutral charge state.

\section{Conclusions}

In this work we propose a first-principles method to compute elastic scattering cross sections of electrons impinging on polycentric systems, such as water and water aggregates. Our approach does not require approximations, beyond the chosen many-body method (that is systematically improvable), on the channels involved in the process. Indeed, the latter are inherently included in the formalism by solving the LS equations for the scattering wave function. Furthermore, the projection of the potential onto a functional space spanned by HGFs allows one to deal with polycentric systems of increasing complexity and different charge states, and to achieve a computational cost similar to that of bound state calculations. \\ 
\indent The modelling of the electron elastic scattering in (polar) molecules typically is carried out either by ab initio methods, typically based on the R-matrix approach \citep{faure_2004,gorfinkiel2005electron} or by following the Mott theory. 
The main difference with the Mott theory is that in our scheme the scattering potential is not a mere sum of atomic contributions, and thus fully accounts for the polycentric nature of the target system, including dipole moments and multiple scattering arising from the accurate treatment of the inter- and intra-molecular potential. This is at odds with the Mott theory, where the elastic scattering by different atoms belonging to the same molecule is considered independently and the total cross section is obtained as a weighted sum of the atomic cross sections. We have demonstrated that both the inclusion of re-scattering and the accurate calculation of the inter- and intra-molecular potentials are crucial to the assessment of the elastic cross section.\\
\indent On the other hand, ab initio methods so far used to calculate the ESCS of electrons do not account for relativistic effects and are typically limited to the modelling of a single water molecule. At variance, our scheme clearly shows how the consideration of the liquid phase nature of water emerges as a significant deviation from the single molecule case; moreover, it is fully relativistic, so important effects present in the elastic scattering of unpolarized electrons, such as spin polarization and flip, are taken fully into account. 
While we recognize that relativistic effects are not of primary importance in water systems, our framework paves the way towards a full understanding  and accurate treatment of the electron scattering also from heavy-element nanoparticles (NPs), such as gold clusters, where the inclusion of relativistic effects is of paramount importance. For example, gold NPs are particularly interesting in hadrontherapy for cancer cure to increase the relative biological effectiveness of ion and electron beams. Moreover, 
the use of polarized electron beams is becoming increasingly important in quantum information theory, which must deal with the decoherence of qubits from quantum noise to protect information. 
Only a quantum-relativistic approach
to the elastic scattering of electron beams by atoms and molecules makes it possible to determine the polarization degree upon scattering and to deliver an accurate account of experimentally produced polarized ensembles of electrons.

\section*{Conflict of Interest Statement}

The authors declare that the research was conducted in the absence of any commercial or financial relationships that could be construed as a potential conflict of interest.

\section*{Author Contributions}

Conceptualization, S.T., S.S. and M.T.; methodology, S.T., S.S., F.T., and M.T.; software, S.S., M.T.; validation, S.S, F.T, and M.T.; investigation, S.S., M.T, and F.T.; resources, S.T., and S.S.; writing--original draft preparation, S.T, T.M., F.T.; writing--review and editing, S.T., S.S., F.T., and M.T.; visualization, M.T., F.T., and S.S.; supervision, S.T, S.S.; project administration, S.T, S.S.; funding acquisition, S.T, S.S. All authors have read and agreed to the published version of the manuscript.

\section*{Funding}
This action has received funding from the European Union under grant agreement n. 101046651.

\section*{Acknowledgments}

We acknowledge fruitful discussions with Drs. M. Dapor, P. De Vera, and Profs. I. Abril and R. Garcia-Molina.

\section*{Data Availability Statement}
The data that support the findings of this study are available from the corresponding author upon reasonable request.

\section*{Appendix}\label{appendice}
The many-particle Dirac equation for particles of mass $m$, interacting via a generic scalar/vector potential is the following:
\begin{equation}\bigg\{ \sum_i(c\alpha_i\mathbf{p}_i+\beta_imc^2+V_i)+\sum_{i<j}[\beta_i\beta_jg_{S,ij}+(1-\alpha_i \cdot \alpha_j)g_{V,ij}]\bigg\}\psi(\mathbf{r}1,...,\mathbf{r}_N)=E\psi(\mathbf{r}_1,...,\mathbf{r}_N),\end{equation}
where $V_i$ is the external potential, $\alpha_i$ and $\beta_i$ are the following $4\times 4$ Hermitian matrices 
\begin{equation}\label{alpha}
\alpha= 
\begin{pmatrix}
0 &  \sigma \\
\sigma & 0 \\
\end{pmatrix}, \hspace{1cm} \beta= 
\begin{pmatrix}
I &  0 \\
0 & -I \\
\end{pmatrix},
\end{equation}
with $I$ the $2 \times 2$ identity matrix and $\sigma$ the Pauli matrices; $g_{V}$ and $g_S$ are respectively the vector and scalar potentials. In second quantization, we can use the Hartree-Fock (HF) approximation to the electron exchange-correlation interaction by assuming:
\begin{displaymath}\biggl\langle \hat\Psi^{\dagger}_{s_1}(\mathbf{r}) \hat\Psi^{\dagger}_{s^\prime_1}(\mathbf{r^\prime}) \hat\Psi_{s^\prime_2}(\mathbf{r^\prime})\hat\Psi_{s_2}(\mathbf{r}) \biggr\rangle=\biggl\langle \hat\Psi^{\dagger}_{s_1}(\mathbf{r}) \hat\Psi_{s_2}(\mathbf{r})\biggr\rangle \biggl\langle \hat\Psi^{\dagger}_{s^\prime_1}(\mathbf{r^\prime}) \hat\Psi_{s^\prime_2}(\mathbf{r^\prime}) \biggr\rangle-\end{displaymath} \begin{equation}\biggl\langle \hat\Psi^{\dagger}_{s_1}(\mathbf{r}) \hat\Psi_{s^\prime_2}(\mathbf{r^\prime})\biggr\rangle \biggl\langle \hat\Psi^{\dagger}_{s^\prime_1}(\mathbf{r^\prime}) \hat\Psi_{s_2}(\mathbf{r}) \biggr\rangle,\end{equation} where the density matrix \begin{equation}\rho_{s^{\prime}_2,s_1}=\biggl\langle \hat\Psi^{\dagger}_{s_1}(\mathbf{r}) \hat\Psi_{s^{\prime}_2}(\mathbf{r^{\prime}})\biggr\rangle\end{equation} is the bloc matrix
\begin{equation}
\rho(\mathbf{r^{\prime}},\mathbf{r})= 
\begin{pmatrix}
\rho_{LL}(\mathbf{r^{\prime}},\mathbf{r}) &  \rho_{LS}(\mathbf{r^{\prime}},\mathbf{r}) \\
\rho_{SL}(\mathbf{r^{\prime}},\mathbf{r})& \rho_{SS}(\mathbf{r^{\prime}},\mathbf{r}) \\
\end{pmatrix},
\end{equation} 
in which LS denote the large and small component of the Dirac bispinor, while $s_1,s_2,s^{\prime}_1$ and $s^{\prime}_2$ label the upper and lower bispinor components. For a many-body systems of electrons interacting only via Coulomb repulsion $V$, the DHF equation can be rewritten as \citep{taioli_2021,de2022simulating}:
\begin{equation}
\begin{pmatrix}
mc^2+V-E &  -c\mathbf{\sigma}\cdot  i \nabla  \\
c\mathbf{\sigma}\cdot  i\nabla & -mc^2-V-E  
\end{pmatrix}
\begin{pmatrix}
\Psi_L\\
\Psi_S
\end{pmatrix}= 0.
\end{equation}
where $V=V_H+V_{exc}$ is written in HF approximation as the sum of the Hartree potential and the exchange interaction. Please notice the different sign in front of the interaction term acting on the upper and lower parts of the spinor.
This matrix equation was used to calculate the electronic structure, and in particular to obtain the relativistic interaction potential of water aggregates.

\bibliographystyle{Frontiers-Harvard} 

\end{document}